\documentclass[%
 reprint,
nofootinbib,
 amsmath,amssymb,
 aps,
prd,
floatfix
]{revtex4-1}

\usepackage{lmodern}
\usepackage{graphicx}% Include figure files
\usepackage{bm}% bold math
\usepackage{hyperref}% add hypertext capabilities
\hypersetup{linktocpage,colorlinks,citecolor={blue},pdfdisplaydoctitle=true,pdfpagemode=UseOutlines,bookmarksnumbered=true}
\usepackage{mathrsfs,dsfont}
\usepackage{color}
\usepackage{cleveref}

\newcommand{\bra}[1]{\langle #1 |} 
\newcommand{\ket}[1]{| #1 \rangle } 
\newcommand{\upd}{\mathrm{d}}
\newcommand{\tr}{\mathrm{tr}}
\newcommand{\ie}[0]{\textit{i.e.} }
\newcommand{\eg}[0]{\textit{e.g.} }

\newcommand{\e}[0]{\mathrm{e}}

\DeclareMathOperator*{\argmin}{argmin}

\begin{document}

\title{Relativistic continuous matrix product states for quantum fields without cutoff}

\author{Antoine Tilloy}
\email{antoine.tilloy@mpq.mpg.de}
\affiliation{Max-Planck-Institut f\"ur Quantenoptik, Hans-Kopfermann-Stra{\ss}e 1, 85748 Garching, Germany}
\affiliation{Munich Center for Quantum Science and Technology (MCQST), Schellingstr. 4, D-80799 M\"unchen}

\begin{abstract}
\noindent 
I introduce a modification of continuous matrix product states (CMPS) that makes them adapted to relativistic quantum field theories (QFT). These relativistic CMPS can be used to solve genuine $1+1$ dimensional QFT without UV cutoff and  directly in the thermodynamic limit. The main idea is to work directly in the basis that diagonalizes the free part of the model considered, which allows to fit its short distance behavior exactly. This makes computations slightly less trivial than with standard CMPS. However, they remain feasible and I present all the steps needed for the optimization. The asymptotic cost as a function of the bond dimension remains the same as for standard CMPS. I illustrate the method on the self-interacting scalar field, a.k.a. \!the $\phi^4_2$ model. Aside from providing unequaled precision in the continuum, the numerical results obtained are truly variational, and thus provide rigorous energy upper bounds.
\end{abstract}

\maketitle

\section{Introduction}
\noindent
Tensor network states (TNS) have become the tool of choice for many-body quantum systems on the lattice \cite{cirac2009-rev,evenbly2014-rev}. They provide a sparsely parameterized manifold of wavefunctions that is well suited for the approximation of the low energy states of local Hamiltonians \cite{molnar2015,hastings2006}. The \emph{bond dimension} $D$ controls the size of the manifold and thus the expressiveness of the class of states. In $1+1$ spacetime dimensions, matrix product states (MPS) \cite{fannes1992} are the simplest instance of TNS. They underpin the earlier density matrix renormalization group  \cite{white1992,white1993,schollwock2005} and provide one of the most efficient numerical methods to study quantum chains. Their extension for $2+1$ dimensional problems, the projected entangled pair states (PEPS) \cite{verstraete2004}, while more difficult to optimize, have been used successfully in a wide range of non-trivial instances including the Hubbard model. Aside from their numerical use, tensor network states are powerful theoretical tools \cite{cirac2020matrix}, \eg for the classification of topological \cite{schuch2010,bultinck2017} and symmetry protected \cite{pollmann2010,chen2011,schuch2011,chen2013} phases of matter.

In light of the progress they enabled in the discrete, it is tempting to use TNS to solve problems in the continuum, and attack quantum field theories (QFTs). This can be done in two ways, (i) by discretizing the QFT first to then use the standard lattice TNS toolbox and finally extrapolate the results (ii) by taking the continuum limit of TNS first to apply them to the continuum model without extrapolation needed. The first option has already provided numerically accurate results in many quantum field theories \cite{milsted2013,banuls2013,banuls2016,banuls2017,kadoh2019,delcamp2020}. However, working directly in the continuum offers crucial advantages: spatial symmetries are preserved and the perilous continuum extrapolations of the results are avoided. It is this neater second option, which consists in defining a manifold of states directly in the continuum, that I am interested in here.

In 2010, Verstraete and Cirac introduced continuous matrix product states (CMPSs) \cite{verstraete2010}, which are a proper continuum limit of MPS particularly adapted to $1+1$ dimensional non-relativistic field theories (like the Lieb-Liniger model). The corresponding numerical toolbox was then developed and extended in the past ten years \cite{ganahl2017,ganahl2018,tuybens2020variational}. More recently, an extension to $d+1$ ($d\geq 2$) dimensions pushing PEPS to the continuum, was put forward in~\cite{tilloy2019}, but without a general numerical toolbox yet. 

The previous continuum ansatz are adapted to non-relativistic theories only, and cannot be used for relativistic QFTs without an additional momentum cutoff $\Lambda$. There are many ways to understand this limitation which I summarize in sec. \ref{sec:UVproblems}. For relativistic QFTs, this partially defeats the purpose of going to the continuum in the first place, since one still needs to extrapolate the final results as $\Lambda \rightarrow + \infty$. This is particularly disappointing given that, at least in $1+1$ dimensions, relativistic QFTs are straightforward to define ``all the way down'', without any cutoff~\cite{glimm1987,fernandez1992}.

My main objective in this paper is to introduce a modification of CMPS, the relativistic CMPS (RCMPS), that is adapted to relativistic QFT in $1+1$ dimension, and does not require any additional cutoff, UV or IR. In a nutshell it is a state $\ket{Q,R}$ parameterized by two $D\times D$ complex matrices $Q,R$ and defined as
\begin{equation}
    \ket{Q,R} = \tr\left\{\mathcal{P} \exp\left[\int \upd x \, Q\otimes \mathds{1} + R\otimes a^\dagger(x)\right]\right\}\ket{0}_a.
\end{equation}
I will explain this formula in more detail later. Let me just mention here that $a^\dagger(x)$ is the Fourier transform of the momentum creation operator $a_k^\dagger$ diagonalizing the free part of the relativistic Hamiltonian under consideration and $\ket{0}_a$ the associated Fock vacuum. Crucially, $a^\dagger(x)$ is \emph{not} the creation operator $\psi^\dagger(x)$, local in the canonically conjugated fields, and used for non-relativistic CMPS. Using $a$ instead of $\psi$ is the only difference between CMPS and RCMPS. We will see later why using $a$ is a good idea and the consequences this choice has. 

I can mention one advantage already. Because RCMPSs allow to work directly with the true theory without cutoff or extrapolation, the results one obtains are genuinely variational, and thus provide rigorous energy upper bounds. Further, while results based on a discretization or with a finite UV cutoff have a limited validity at high momenta, the RCMPS are exact in the UV. 

The price to pay is that RCMPS are obtained from CMPS via change of basis $\psi,\psi^\dagger \rightarrow a,a^\dagger$ that is not local. In the new basis, the Hamiltonian density of a relativistic QFT is no longer strictly local, and only exponentially decaying. In practice, the lack of strict locality of the Hamiltonian density introduces minor complications in the computations. However, and perhaps surprisingly, these complications do not change the asymptotic scaling of the cost compared to regular CMPS, which remains $\propto D^3$. Further, the energy density seems to converge fast as a function of $D$ as one would expect \cite{huang2015computing}, while the cost which means RCMPS provide an efficient class of states for relativistic QFTs.

In addition to their direct use for relativistic QFT in $1+1$ dimensions, demonstrated in this paper, RCMPS could be used as an auxiliary step to solve non-relativistic QFTs in $2+1$ dimensions. Indeed, continuous generalizations of PEPS currently lack a general purpose optimization algorithm. The missing routine is a function to solve a \emph{relativistic} QFT in one dimension less, that is in $1+1$ space-time dimensions \cite{tilloy2019}. The relativistic nature of this auxiliary theory comes from the need for Euclidean invariance in the $2$ space dimensions of the original non-relativistic theory \cite{tilloy2019}. For that task, RCMPS would fit the bill better than CMPS and preserve exact Euclidean invariance at short distances.

The present paper is structured as follow. I first discuss standard CMPS and their limitations for relativistic theories in sec.~\ref{sec:cmps} before introducing and discussing the RCMPS ansatz in sec.~\ref{sec:rcmps}. I first explain how to compute expectation values in sec.~\ref{sec:computations} and then show how to optimize the state to find  ground states in sec.~\ref{sec:optimization}. The numerical results for the ground energy and correlation functions, which result from this optimization, are presented for the $\phi^4$ model in sec.~\ref{sec:application}. I finally discuss some extensions in sec.~\ref{sec:extensions}, namely a slight variation of RCMPS with more general basis, and explain how one could obtain more general observables by extending known CMPS techniques. I conclude with a general discussion of the method in sec.~\ref{sec:discussion}, comparing it with renormalized Hamiltonian truncation (RHT). A quicker presentation of the results, emphasizing the importance and difficulty of the variational method, is presented in a companion letter \cite{rcmps_letter}.

\section{Standard continuous matrix product states}\label{sec:cmps}
\subsection{Definition and main properties}
\noindent
Continuous matrix product states (CMPS) were introduced by Verstraete and Cirac \cite{verstraete2010} in 2010. They extend the successful matrix product states ansatz from the lattice to the continuum. Concretely, for a bosonic quantum field theory on the closed line $[-L,L]$ with (non-relativistic) creation and annihilation operators $(\psi^\dagger,\psi)$, a CMPS is the state
\begin{equation}
    \ket{Q,R}_\mathrm{nr}= \tr\left\{\mathcal{P} \exp\!
    \left[\int_{-L}^L \!\upd x \, Q\otimes \mathds{1} + R\otimes \psi^\dagger(x)\right]\right\}\ket{0}_\psi\, ,
\end{equation}
where by definition $[\psi(x),\psi^\dagger(y)] = \delta(x-y)$, $\mathcal{P}$ is the path ordering operator, and $\ket{0}_\psi$ is the Fock vacuum annihilated by all the $\psi(x)$. The state is parameterized by two $D\times D$ matrices $Q$ and $R$, where $D$ is the bond dimension, and which contain all the variational freedom. The trace is taken over the associated $D$ dimensional Hilbert space, and implements periodic boundary conditions (more general boundary conditions are possible by inserting a matrix $B$ in the trace).

CMPS can be derived from a genuine continuum limit of MPS and thus share many of their properties \cite{haegeman2013}. In particular, all the correlation functions of local operators can be computed explicitly and efficiently. This is seen by introducing the generating functional $\mathcal{Z}_{j',j}$:
\begin{equation}\label{eq:generatingfunction}
\mathcal{Z}_{j',j}=\frac{\bra{Q,R} \exp\left(\int j'\, \psi^\dagger \right) \exp\left(\int j\, \psi  \right) \ket{Q,R}_\mathrm{nr}}{\langle Q,R|Q,R\rangle_\text{nr}} \,,
\end{equation}
which can be used to compute all normal-ordered correlation $N$-point functions of local operators, \eg
\begin{align}
\label{eq:correlexample}
\langle\psi^\dagger(x) \psi(y)\rangle:=& \frac{\langle Q,R | \psi^\dagger(x) \psi(y)| Q,R\rangle_\mathrm{nr}}{\langle Q,R | Q,R\rangle_\textrm{nr}} \nonumber \\
=& \frac{\delta}{\delta j'(x)}\frac{\delta}{\delta j(y)} \mathcal{Z}_{j',j} \bigg|_{j,j'=0}\,. 
\end{align}
Using Wick's theorem, one can show \cite{haegeman2013} that this generating functional has an exact expression:
\begin{equation}
    \mathcal{Z}_{j',j} \! = \! \tr\left\{\mathcal{P} \exp\!
    \bigg[\int_{-L}^L \!\!\!\upd x \, \mathbb{T}  + j(x)  R \otimes \mathds{1} +j'(x) \mathds{1}\otimes \bar{R}\bigg]\right\}\,
\end{equation}
where $\mathbb{T} = Q\otimes \mathds{1} + \mathds{1}\otimes \bar{Q} + R\otimes \bar{R}$ is the transfer operator and the trace is now taken over the tensor product of two copies of the original $D$ dimensional Hilbert space. For example, for the simple two-point function of eq. \eqref{eq:correlexample}, using the exact expression gives, for $x\geq y$:
\begin{equation}\label{eq:twopointexplicit}
    \langle\psi^\dagger(x) \psi(y) \rangle \!=\! \tr\Big[ \e^{(L-x)\mathbb{T}} (\mathds{1}\otimes \bar{R})
\e^{(x-y)\mathbb{T}} (R\otimes\mathds{1}) \e^{(y+L)\mathbb{T}}\Big].
\end{equation}
The state $\ket{Q,R}_\mathrm{nr}$ is parameterized in a redundant way, in the sense that different $Q,R$ pairs give the same state. In particular, conjugating both matrices with an invertible matrix $U$ keeps the state invariant. This freedom can be used to choose a gauge in which only the anti-Hermitian part $K$ of $Q$ is free \cite{haegeman2010}:
\begin{equation}\label{eq:left_gauge}
    Q := -iK -\frac{1}{2} R^\dagger R\, .
\end{equation}
In this gauge, the transfer operator $\mathbb{T}$ is of the Lindblad form, hence negative, and generically has a single largest eigenvalue $\lambda_0=0$ with associated left and right eigenvectors $\bra{\ell_0}$ and $\ket{r_0}$. Using this form provides convenient expressions in the thermodynamic limit, as $\e^{L\mathbb{T}} \rightarrow \ket{r_0}\bra{\ell_0}$ when $L\rightarrow + \infty$. For example, eq. \eqref{eq:twopointexplicit} simplifies to:
\begin{equation}\label{eq:twopoint_thermo}
     \langle\psi^\dagger(x) \psi(y) \rangle \!=\! \bra{\ell_0} (\mathds{1}\otimes \bar{R})
\e^{(x-y)\mathbb{T}} (R\otimes\mathds{1}) \ket{r_0}\, .
\end{equation}
In what follows, and in particular for the upcoming relativistic extension, I will always work directly in the thermodynamic limit.

A CMPS is typically used to find the ground state of a non-relativistic QFT. The archetypal example is the Lieb-Liniger model with Hamiltonian $H_\text{LL}$:
\begin{align} \label{eq:LL_hamiltonian}
    H_\text{LL} 
    &=\int_\mathbb{R}  \partial_x \psi^\dagger\partial_x\psi + c\,  \psi^\dagger\psi^\dagger\psi\psi - \mu \psi^\dagger\psi \\
    &= \int_\mathbb{R}  h_\text{LL} \; . 
\end{align}
With the generating functional, it is possible to compute the expectation value of the $3$ terms of the Hamiltonian density $\langle h_\text{LL}\rangle_{Q,R} = f(Q,R)$, where $f$ is a simple function containing expectation values of products of $Q$ and $R$ as in eq. \eqref{eq:twopoint_thermo}. One then minimizes the energy density over the matrices $Q$ (or $K$) and $R$ to get close to the ground state:
\begin{equation}
    \ket{\mathsf{ground}}\simeq \ket{Q,R} \;\;\text{where}\;\; Q,R = \argmin \, \langle h_\text{LL}\rangle_{Q,R}  .
\end{equation}
The approximation can get \emph{arbitrarily} good as the bond dimension (and thus the size of the variational manifold) is increased because CMPS are dense in the Fock space \cite{haegeman2013}. Further since the method is variational, we also know that the energy we find for finite $D$ always upper bounds the true ground energy
\begin{equation}
    \varepsilon_0 := \langle h_\text{LL}\rangle_{\textsf{ground}} \leq \langle h_\text{LL}\rangle_{Q,R} \,  .
\end{equation}
In practice, one can simply feed the expression of the energy density as a function of $Q$ and $R$ to a standard numerical minimizer as was done originally in \cite{verstraete2010}. However it is typically much more efficient, especially for large $D$, to use more elaborate tangent space methods \cite{vanderstraeten2019tangentspace}. I explain how they work in sec. \ref{sec:optimization}. In a nutshell, the latter require computing the exact gradient with respect to the variational parameters as well as the natural metric on the tangent space of CMPS induced by the real part of the Hilbert space scalar product. The optimization is then done through gradient descent on the corresponding differentiable manifold. At this stage, all that matters for us is that it can be done and that it is efficient: one can easily optimize CMPSs of large bond dimensions~\cite{ganahl2017} (without barren plateaus or other pathology).

\subsection{UV problems in relativistic theories}\label{sec:UVproblems}
\noindent
CMPS are well suited for non-relativistic theories, which they approximate well ``all the way down'', without short distance cut-off. However, once we move to relativistic QFT, it becomes necessary to introduce a UV cut-off. This is best understood on the free boson, which was discussed in the CMPS context in~\cite{stojevic2015}. The free boson Hamiltonian is
\begin{equation}
    H_\text{fb}=\frac{1}{2}\int_{\mathbb{R}} \pi^2+ (\partial_x \phi)^2 + m^2 \phi^2 \,,
\end{equation}
where $m$ is the mass and $\pi,\phi$ are canonically conjugated $[\phi(x),\pi(y)]=i\delta(x-y)$. To deal with such a Hamiltonian with a CMPS, it is tempting to express the field operator and its conjugate as a function of a non-relativistic creation-annihilation pair $\psi^\dagger,\psi$:
\begin{align}
    \phi &= \sqrt{\frac{1}{2\Lambda}}(\psi+\psi^\dagger) \, ,\\
    \pi&= \sqrt{\frac{\Lambda}{2}}\; (\psi-\psi^\dagger) \, ,
\end{align}
which introduces a new arbitrary mass scale $\Lambda$. In this basis, the Hamiltonian $H_\text{fb}$ is still the integral of a local density, and thus it is straightforward to evaluate the energy density on a CMPS $\ket{Q,R}_\text{nr}$.

This unfortunately leads to mild and serious divergences. The first mild one is that $H_\text{fb}$ is a priori not normal ordered in the $\psi^\dagger,\psi$ operator basis. This is simply solved by considering $:H_\text{fb}{:}_\mathrm{\psi}$ instead. The serious divergence comes from the fact that $:H_\text{fb}{:}_\mathrm{\psi}$ not only contains the standard non-relativistic kinetic energy $\partial_x\psi^\dagger\partial_x\psi$ but also $\partial_x\psi\partial_x\psi + \text{h.c.}$. The latter is divergent when evaluated on a generic CMPS \cite{haegeman2010}. This second divergence can be cured by adding an adapted UV regulator to the Hamiltonian and considering \cite{stojevic2015}
\begin{equation}
    H^\Lambda_\text{fb}=\frac{1}{2}\int_{\mathbb{R}} \pi^2+ (\partial_x \phi)^2 + m^2 \phi^2 + \underset{ \text{regulator}}{\underbrace{\frac{1}{\Lambda^2} \left(\partial_x \pi\right)^2}}\, ,
\end{equation}
which kills the problematic terms. To reach a fixed precision, one then needs to increase the bond dimension D as the cutoff $\Lambda$ is lifted. Similarly for fermionic theories, it was observed in \cite{haegeman2010-relativistic} that one needs to add a UV cutoff to the Hamiltonian, this time not to ensure the finiteness of the results, but to make the optimization problem well behaved. In a nutshell, without a cut-off, the CMPS reduces the energy density by approximating larger and larger momentum modes as the optimization proceeds, completely missing the IR which contributes to the energy only in a subleading way. Independently of CMPS, this sensitivity to high frequencies in variational methods was considered by Feynman as one of the main difficulties preventing its use in relativistic QFT \cite{feynman1988}.

Zooming out from the CMPS peculiarities, this situation is not surprising, since we are working in the wrong operator basis. Indeed, the operator $:H_\text{fb}{:}_\psi$ is not bounded from below with this specific ordering. Hence even if we manage to have a finite energy density when computing CMPS expectation values, we are always infinitely far from the true ground state. Further, the free boson is a conformal field theory (CFT) at short distances (the massless free boson). It is thus no surprise that a CMPS, which is adapted to systems with a gap and exponentially decaying correlation functions, completely fails to capture the short distance behavior of relativistic models. Yet another way to see the problem is that the ground state of the free boson has an infinite density of non-relativistic particles, whereas a CMPS always gives a finite density. 

Note that these UV problems are not made easier if one adds a relevant interaction, as the latter becomes negligible at short distances. The relativistic continuous matrix product state ansatz will solve these short distance problems already present in the free theory, but also allow to deal with interactions without additional difficulty.

\section{Relativistic continuous matrix product states} \label{sec:rcmps}

\subsection{Intuition}
\noindent
In the standard textbook approach to quantum field theory \cite{peskin1995}, the problem of infinite particle density is solved at the very beginning by changing of basis (and in fact even of Hilbert space). Typically, one expands the field operator in new creation-annihilation modes that diagonalize the Hamiltonian:
\begin{align}\label{eq:modeexpansion}
    \phi(x) &= \frac{1}{2\pi} \int \upd k \sqrt{\frac{1}{2 \, \omega_k}} \left(\e^{ikx} a_k + \e^{-ikx} a^\dagger_k \right) \\
        \pi(x) &= \frac{1}{2\pi i} \int \upd k \sqrt{\frac{\omega_k}{2}} \left(\e^{ikx} a_k - \e^{-ikx} a^\dagger_k \right) \, ,
\end{align}
where $\omega_k=\sqrt{k^2 + m^2}$ and $[a_k,a_{k'}^\dagger]=2\pi\delta(k-k')$.
In condensed matter physics, this would be called a Bogoliubov transform. In this new basis, the Hamiltonian is diagonal
\begin{equation}
    H_\text{fb} = \frac{1}{2\pi}\int_\mathbb{R} \upd k  \,  \omega_k \; \frac{a^\dagger_k a_k + a_k a^\dagger_k}{2} \, .
\end{equation}
Then, normal ordering $H_\text{fb} \rightarrow :H_\text{fb}{:}_a$ removes the infinite vacuum energy contribution and one finds that the ground state is simply the Fock vacuum annihilated by all the $a_k$'s: $\ket{\textsf{ground}} = \ket{0}_a$. The Hamiltonian $:H_\text{fb}{:}_a$ further is a well defined operator on the Fock space generated by creating relativistic particles from the vacuum.

The crucial observation is that in $1+1$ dimensions, this normal ordering procedure is typically\footnote{Normal ordering is sufficient to make polynomials in the field well defined. For more general potentials, like $\cos(b\phi)$, normal ordering is sufficient only for $b$ small enough.} sufficient to cure all the divergences that can appear, even when adding interactions. In particular, the $\phi^4$ Hamiltonian $H$ which we will consider in more detail later,
\begin{equation}\label{eq:phi4split}
    H = :H_\text{fb}{:}_a + g \int_\mathbb{R} :\phi^4{:}_a\, ,
\end{equation}
is a perfectly legitimate and regular Hamiltonian on the free Fock space. From now on, I will omit the subscript on the normal ordering which will always be done with respect to $a,a^\dagger$ unless otherwise stated.

The Fock space generated by acting with $a^\dagger_k$ on $\ket{0}_a$ is much more adapted to relativistic theories than the Fock space generated by acting with $\psi^\dagger$ on $\ket{0}_\psi$, because the former solves the scale invariant short distance behavior of the theory exactly with its vacuum. 
However, the operators $a_k,a_k^\dagger$ are not adapted to CMPS because the latter encode the state in real space, and not momentum space. Further, the Hamiltonian $H$ is not translation invariant in momentum, which is the situation CMPS are most adapted for. A tempting workaround is to simply Fourier transform $a_k$ to get $a(x)$ 
\begin{align}\label{eq:fourier}
    a(x) &= \frac{1}{2\pi}\int \upd k \, \e^{ikx} a_k \, ,
\end{align}
which verifies $[a(x),a^\dagger(y)] = \delta(x-y)$. Note the crucial fact that $\psi(x) \neq a(x)$, as the Fourier transform \eqref{eq:fourier} does not contain the factors $\omega_k$.  Intuitively, $a^\dagger(x)\ket{0}_a$ corresponds to a (bare) relativistic particle localized in $x$. Working with this new operator basis is the key to make CMPS adapted to relativistic theories.

\subsection{Definition and basic properties}

\noindent
The previous discussion naturally leads us to introduce the relativistic CMPS (RCMPS) ansatz:
\begin{equation}
    \ket{Q,R} = \tr\left\{\mathcal{P} \exp\left[\int \upd x \, Q\otimes \mathds{1} + R\otimes a^\dagger(x)\right]\right\}\ket{0}_a
\end{equation}
where $a^\dagger(x)$ is the Fourier transform of the creation operator diagonalizing the free part of the relativistic Hamiltonian under consideration. Note that we choose to work directly in the thermodynamic limit, and thus the state $\ket{Q,R}$ has no cut-off, UV or IR (although the latter is trivial to reintroduce). 

The properties of the state are the same as before if one replaces $\psi$ by $a$, as these operators obey the same algebra. A large part of the theory of CMPS can thus be reused. In particular, all normal-ordered $N$-point correlation functions of $a,a^\dagger$ can be computed efficiently using the same generating functional. Additionally, we have that RCMPS are dense in the appropriate QFT Fock space (the manifold is maximally expressive) and thus that the precision can be arbitrarily refined by increasing $D$.

\subsection{Consequence on the Hamiltonian density} \label{sec:consequences}
\noindent
The main difference with the non-relativistic case is that, once written as a function of $a(x)$, the Hamiltonian of a (massive) relativistic theory is not strictly local, but only exponentially decaying. Indeed a relativistic Hamiltonian is local in the field $\phi(x)$ and its conjugate, which do not have a local expression as a function of $a(x)$. More precisely, 
\begin{align}\label{eq:convolution}
    \phi(x) &= \frac{1}{2\pi} \int \upd k \sqrt{\frac{1}{2 \, \omega_k}} \left(\e^{ikx} a_k + \e^{-ikx} a^\dagger_k \right) \nonumber \\
     &= \frac{1}{2\pi} \int  \frac{\upd k \, \upd y}{\sqrt{2 \, \omega_k}} \left(\e^{ik(x-y)} a(y) + \e^{-ik(x-y)} a^\dagger(y) \right) \nonumber\\
     &=\int \upd y\;  J(x-y) \left[a(y) + a^\dagger(y) \right]\,
\end{align}
where $J(x)$ is a smooth kernel for $x\neq 0$ (and not a Dirac distribution) which I will make explicit later.

Consequently, the Hamiltonian density of a relativistic theory with terms of order up to $n$ in the field $\phi$ can be written as a function of $a(x)$ with $n$ integrals, a bit informally:
\begin{align}
    H &= \int \upd x\, h(x) \\
    &=\int \! \upd x \int \!\upd x_1 .. . \upd x_n K(x_1,...,x_n) a^{(\dagger)}(x_1) ... a^{(\dagger)}(x_n) \,
\end{align}
where $a^{(\dagger)}$ is a compact notation to convey the fact that both creation and annihilation occur in a sum, and $K$ is a kernel that decays exponentially as a function of the difference of its arguments.

Is this non-locality a problem? Technically, it certainly induces complications in evaluating the energy density: we have exact expressions for $N$ point functions of $a,a^\dagger$ which we then have to integrate over some kernel (instead of taking them at equal point for a local density). This is however not insurmountable, and as I will show in the next section, one can still compute the Hamiltonian density at a cost $\propto D^3$. 

More crucially, and provided we can optimize them, do we expect the expressive power of RCMPS to be lower for such Hamiltonians? There is no reason to think so, and in fact CMPS have already be applied to Hamiltonians with exponentially decaying interactions in the non-relativistic context \cite{rincon2015}. Intuitively, we are merely introducing a new length scale $m^{-1}$ that replaces the lattice scale in lattice models. This is more physical than introducing a much smaller and arbitrary cut-off scale $\Lambda^{-1} \ll m^{-1}$ as was done previously. Other choices of length-scales that could further improve precision are discussed in sec. \ref{sec:extensions}.

\section{Efficiently computing RCMPS expectation values} \label{sec:computations}

\subsection{Naive direct evaluation}
\noindent The operators $a^\dagger(x),a(y)$ have the same commutation relations as the $\psi^\dagger(x),\psi(y)$ of standard non-relativistic CMPS and thus all the formulas follow. In particular, one can straightforwardly compute the exact expectation values of normal-ordered products of $a^\dagger$ and $a$ using the generating functional \eqref{eq:generatingfunction}. 

Obtaining simple functions of the field $\phi$, \eg expectation values of normal-ordered monomials $\langle :\! \phi^n\!: \rangle$ is less straightforward because the field $\phi$ is obtained from $a,a^\dagger$ with a convolution \eqref{eq:convolution}. Hence, to compute the expectation value of $\langle :\! \phi^n\!: \rangle$, one \emph{a priori} needs to compute $n$ integrals (in fact $n-1$ using translation invariance) of exact $a,a^\dagger$ correlation functions $\langle a^{(\dagger)}(x_1) a^{(\dagger)}(x_2) \cdots a^{(\dagger)}(x_n) \rangle$. This is feasible for low bond dimension and can be used as a sanity check, but is prohibitively expensive for a full optimization of the state\footnote{This was, unfortunately, the strategy I first followed.}.

\subsection{Vertex operators}
\noindent For efficient computations of functions of the field $\phi$, the starting point is to compute expectation values of vertex operators:
\begin{equation}\label{eq:vertex_definition}
    \langle V_b\rangle:=\bra{Q,R} :\! \e^{b\phi(x)}\!: \ket{Q,R}\, ,
\end{equation}
which we may evaluate in $x=0$ without loss of generality because of translation invariance.
Such an expectation values seems even more difficult to compute than field monomials at first sight. Indeed, using the naive approach above and expanding the exponential \eqref{eq:vertex_definition} provides $n$ integrals at each order $n$. However, it turns out one can directly compute the expectation value without expanding the exponential.

To this end, we simply express the vertex operators as a function of $a(x)$
\begin{align}
    :\! \e^{b\phi(0)}\!: &= :\!\exp\left[ \frac{b}{2\pi }\int \upd x \! \int \frac{\upd k}{\sqrt{2\omega_k}} \e^{-ikx} a(x) + \e^{ikx} a^\dagger(x) \right]\!\! :\nonumber\\
    &= \exp\left[b \int \upd x \, J(x) a^\dagger (x) \right]\exp\left[b \int \upd x \, J(x) a (x)\right] \label{eq:ordered_vertex}
\end{align}
where $J$ is a real function
\begin{align}
    J(x):&= \frac{1}{2\pi} \int \frac{\upd k}{\sqrt{2\omega_k}} \, \e^{-ikx} \label{eq:J_def}\\
    &= \frac{K_{1/4}(|x/m|)}{2^{9/4}\sqrt{\pi}\, \Gamma(5/4)\, |x/m|^{1/4}} \,
    \end{align} 
and $K_{\nu}(x)$ (not to be confused with the matrix $K$) is the modified Bessel function of the second kind.
We see from \eqref{eq:ordered_vertex} the crucial fact that the expectation value of a vertex operators is simply the generating functional itself, with $b J$ as source, \ie $\langle V_b\rangle= \mathcal{Z}_{bJ,bJ}$, thus explicitly
\begin{equation}
    \langle V_b\rangle  = \tr \left[\mathcal{P}\exp\int_\mathbb{R} \upd x\,  \mathbb{T} + bJ(x) \left(R \otimes \mathds{1} + \mathds{1} \otimes \bar{R} \right)\right] \,.
\end{equation}
Inside the trace, this is simply the solution of an ordinary differential equation that can be solved numerically. 

To reduce the computational cost, one can use the standard trick of matrix product states which consists in mapping the tensor product Hilbert space $\mathbb{C}^D\otimes \mathbb{C}^D$ to the space of matrices $\rho$ acting on $\mathbb{C}^D$. Introducing the super-operators
\begin{align}
    \mathcal{L}\cdot \rho &= -i[K,\rho] + R\rho R^\dagger - \frac{1}{2} \left(R^\dagger R \rho  + \rho R^\dagger R \right) \label{eq:def_Lindblad}\\
    \mathcal{R}(x) \cdot \rho &= J(x) (R\rho + \rho R^\dagger)
\end{align}
we have
\begin{equation}
    \langle V_b\rangle = \tr\left\{\mathcal{P}\exp\left[\int_\mathbb{R} \upd x \, \mathcal{L} + b \mathcal{R}(x)\right]\cdot \rho_\text{ss}\right\}
\end{equation}
where $\rho_\text{ss}$ is the stationary state of $\mathcal{L}$ normalized with trace $1$. Rewriting the path-ordered exponential as the solution of an ODE we have
\begin{equation}\label{eq:vertex_Lindblad}
    \langle V_b\rangle = \lim_{x\rightarrow +\infty} \tr \left[\rho_x\right]
\end{equation}
where $\lim_{x\rightarrow-\infty} \rho_x = \rho_\text{ss}$ and
\begin{equation}\label{eq:vertex_ode}
    \frac{\upd}{\upd x} \rho_x = \mathcal{L} \cdot \rho_x + b\mathcal{R}(x) \cdot \rho_x
\end{equation}
The limit \eqref{eq:vertex_Lindblad} is well defined because $J(x)$, and thus $\mathcal{R}(x)$, decrease exponentially fast at infinity and are integrable in $0$. Because of this fast decay at infinity, one could in fact use any density matrix as initial state. Using a simple ODE solver, \eg a backward differential formula (BDF) solver, one can obtain the limit in \eqref{eq:vertex_Lindblad} to arbitrary precision with only a reasonable number of subdivisions. The total computational cost is proportional to the cost of applying the super-operator $\mathcal{L} + b\mathcal{R}(x)$ on a density matrix and thus scales $\propto D^3$ only.
\subsection{Field monomials}
\noindent To compute field monomials, one can differentiate vertex operators with respect to their exponent $b$
\begin{equation}
    \langle :\phi^n\!:\rangle= \frac{\partial^n}{\partial b^n} \langle V_b\rangle \bigg|_{b=0} \, .
\end{equation}
This allows to obtain $\langle :\phi^n\!:\rangle$ by directly differentiating the ODE \eqref{eq:vertex_ode}. Doing so yields 
\begin{equation}
    \langle :\phi^n\!:\rangle = \lim_{x\rightarrow + \infty} \tr\left[\rho^{(n)}_x\right]
\end{equation}
where $\rho^{(k)}:= \partial_b^k \rho^b|_{b=0}$ obey $n$ coupled matrix ODE
\begin{equation}
    \frac{\upd}{\upd x} \rho^{(k)}_x = \mathcal{L}\cdot \rho^{(k)}_x + \mathcal{R}(x) \cdot \rho^{(k-1)}_x
\end{equation}
with the convention that $\rho^{(0)}_x\equiv \rho_\text{ss}$ and for $k>0$, $\rho^{(k)}_{-\infty}=0$. Solving the ODE above numerically provides arbitrarily accurate approximations of $\langle :\phi^n\!:\rangle$ at a cost $\propto n\times D^2$.

\subsection{Kinetic term}
\noindent In addition to exponentials and polynomials of the field $\phi$, it is important to be able to compute the expectation value of the free part of the Hamiltonian.

For convenience, we consider directly the Hamiltonian for the \emph{massive} free boson, but since the mass term $m^2\langle :\!\phi^2\!:\rangle$ is also computable, it could be subsequently subtracted to obtain the pure kinetic term. The free boson Hamiltonian can be expressed as a function of the momentum space creation and annihilation operators $a_k^\dagger,a_k$ and reads
\begin{equation}
    :H_\text{fb}: = \frac{1}{2\pi}\int \upd k \, \omega_k \; a^\dagger_k a_k\, .
\end{equation}
The corresponding Hamiltonian density $h_\text{fb}(x)$ is
\begin{equation}
    :h_\mathrm{fb}(x): = \frac{1}{2\pi} \int \upd k \upd y \, \omega_k\;  \e^{ik(y-x)}a^\dagger(y) a(x) \, .
\end{equation}
As before, we would like to write this density as a derivative of a vertex operator. If we try to mirror the reasoning of the previous subsection, we face the issue that the natural source $\tilde{J}$ that appears now is the Fourier transform of $\sqrt{\omega_k}$. This is not a function but only a distribution. To get a true function, we can divide and multiply by $\omega_k^2=m^2+k^2$, and interpret the $k^2$ term on the numerator as a $\partial_x\partial_y$ derivative. This gives 
\begin{equation} 
    :h_\mathrm{fb}(x): = \frac{1}{2\pi} \int \upd y  \, \frac{\upd k}{\omega_k}\;  \e^{ik(y-x)} (m^2 + \partial_y \partial_x ) a^\dagger(y) a(x) \, .
\end{equation}
We are back to an expression that depends on the source $J(x)$ that we introduced before \eqref{eq:J_def}
\begin{equation}
    \begin{split}
    \langle :h_\mathrm{fb}: \rangle =& 2 m^2 \left \langle \int \upd x  J(x) a^\dagger(x)  \!\! \int \upd y J(y) a(y)\right\rangle \\ 
    + & 2 \left\langle \int \upd x J(x) \partial_x a^\dagger(x) \!\! \int \upd y J(y) \partial_y a(y) \! \right\rangle  
\end{split}
\end{equation}
except this time derivatives of $a,a^\dagger$ appear as well. This gives
\begin{align}
  \langle :h_\mathrm{fb}: \rangle  =& 2 \left[m^2 \frac{\partial}{\partial b_1} \frac{\partial}{\partial b_2} \mathcal{Z}_{b_1J,b_2 J}   + \frac{\partial}{\partial b_1} \frac{\partial}{\partial b_2} \mathcal{Y}_{b_1J,b_2 J} \right]_{b_{1,2}=0}
\end{align}
where $\mathcal{Y}_{j',j}$ is the generating functional of normal ordered correlation functions of $\partial_x a^\dagger(x),\partial_y a(y)$. This generating functional also has an exact expression, which is easily derived by differentiating correlation functions obtained from $\mathcal{Z}$ with respect to position
\begin{equation}
    \mathcal{Y}_{j',j} \! = \! \tr\left\{\mathcal{P} \exp\!
    \bigg(\int \! \, \mathbb{T}  + j [Q,R] \otimes \mathds{1}  +j'\mathds{1}\otimes [\bar{Q}, \bar{R}]\bigg)\right\}\, .
\end{equation}
As before, $\frac{\partial}{\partial b_1} \frac{\partial}{\partial b_2} \mathcal{Z}_{b_1J,b_2 J}$ and $\frac{\partial}{\partial b_1} \frac{\partial}{\partial b_2} \mathcal{Y}_{b_1J,b_2 J}$ can be obtained by solving simple ODEs.
To this end, we introduce $\rho_x:=\rho^{b_1 b_2}_x$ with initial condition $\rho_{-\infty}=\rho_\text{ss}$ and dynamics
\begin{equation}
    \frac{\upd}{\upd x} \rho_x = \mathcal{L} \cdot \rho_x + b_1 J(x) R \rho_x +b_2 J(x) \rho_x R^\dagger
\end{equation}
and $\sigma_x:=\sigma^{b_1 b_2}_x$ with initial solution $\sigma_{-\infty} = \rho_\text{ss}$ and dynamics 
\begin{equation}
    \frac{\upd}{\upd x} \sigma_x = \mathcal{L} \cdot \sigma_x + b_1 J(x) [Q,R] \sigma_x +b_2 J(x) \sigma_x [R^\dagger,Q^\dagger]
\end{equation}

We further introduce notations for the partial derivatives  $\rho^{(1,0)}:= \partial_{b_1} \rho|_{b_{1,2}=0}$, $\rho^{(0,1)}:= \partial_{b_2} \rho|_{b_{1,2}=0}$ and $\rho^{(1,1)}:= \partial_{b_1}\partial_{b_2} \rho|_{b_{1,2}=0}$. They obey
\begin{align}
    \frac{\upd}{\upd x} \rho^{(1,0)} &= \mathcal{L}\cdot \rho^{(1,0)}_x + J(x) R \rho_0 \\
      \frac{\upd}{\upd x} \rho^{(0,1)}_x &= \mathcal{L}\cdot \rho^{(0,1)}_x + J(x)  \rho_0 R^\dagger \\
        \frac{\upd}{\upd x} \rho^{(1,1)} &= \mathcal{L}\cdot \rho^{(1,1)}_x + J(x) R \rho_x^{(0,1)} + J(x) \rho_x^{(1,0)} R^\dagger
\end{align}
The same system of ODEs can be obtained for $\sigma$, replacing $R$ by $[Q,R]$ and $R^\dagger$ by $[R^\dagger,Q^\dagger]$. Finally, the expectation value we are looking for is obtained from the trace of the solutions
\begin{equation}
     \langle :h_\mathrm{fb}: \rangle = 2 \lim_{x\rightarrow +\infty}  \tr\left[m^2\rho^{(1,1)}_x + \sigma_x^{(1,1)}\right]\, .
\end{equation}
Hence the expectation value of the massive free boson Hamiltonian density $\langle :h_\mathrm{fb}: \rangle$ can be computed by solving 2 systems of 3 coupled matrix ODEs, and thus can be obtained to arbitrary precision at a cost $\propto D^3$.

\section{Optimization} \label{sec:optimization}

\subsection{Failure of naive optimization}
\noindent
Using the results in the previous section, one obtains an expression for the energy density of the form $\langle h\rangle = f(Q,R)$ where $f$ is a function of the matrices $R$ and $Q$ (in practice $K$) that can be evaluated efficiently on a classical computer at a cost $\propto D^3$. One may thus simply input this function to a standard minimizer, that will typically use a gradient computed by finite differences and hope for the best. This is what was done in the original paper on CMPS applied to the Lieb-Liniger model~\cite{verstraete2010}. For our model, this approach works reasonably well up to $D=4$, after which all the standard optimizers (\eg L-BFGS or conjugate gradient) get stuck in plateaus. To understand why it happens and go beyond such small values of D, we need to do better, and use a tangent space approach.

\subsection{Tangent space approach}
\noindent
It is well known that the notion of steepest descent in optimization depends on a choice of metric. More precisely, if we want to minimize a function $f(x)$ where $x=\{x^\mu\}_{\mu=1}^N$ is a vector of parameters (think of the coefficients of $R,Q$), we can go down the steepest descent direction $u$ defined by 
\begin{equation}
    u=\argmin_{\|u\|= 1} \langle \nabla f, u\rangle_x \, .
\end{equation}
This scalar product on the tangent space $\langle u,v \rangle_x = g_{\mu\nu}(x) u^\mu v^\nu$ and associated metric $g_{\mu\nu}$ are a priori arbitrary. The notion of ``steep'' depends on a metric, and what is steep for the ``right'' metric $g_{\mu\nu}$ may look like a plateau for the naive $\delta_{\mu\nu}$ metric if $g_{\mu\nu}$ is singular. If there are many parameters, the naive metric has no reason to be good.

But what is the right metric? It is one where the distance between parameter values is proportional to how much they change the function one optimizes. An excellent choice of metric is thus given by the Hessian $\text{Hess}_{\mu\nu}:=\partial_\mu\partial_\nu f$ of the function one is optimizing. Taking this metric gives the descent direction $u \propto - [\text{Hess}^{-1}]^{\mu\nu} \partial_\nu f $ where $ [\text{Hess}^{-1}]^{\mu\nu}\text{Hess}_{\nu\rho} =\delta^\mu_\rho$, which corresponds to the famous Newton method. This matrix is costly to estimate for RCMPS, because it requires computing $\propto D^4$ derivatives of the energy density, instead of $\propto D^2$ if we only compute the gradient. 

There is another natural option that comes from the fact that, in our case, the tangent space is also a Hilbert space \cite{hackl2020}. Indeed, let us write $\ket{x} = \ket{Q,R}$ a state in the manifold of RCMPS. Then a natural tangent space metric is simply the Hilbert one 
\begin{equation}
    g_{\mu\nu}(x):=  \text{Re}\left[(\partial_\mu \bra{x})( \partial_\nu \ket{x})\right].
\end{equation}
It provides a notion of distance between parameter values associated to how much they change the quantum state (instead of the energy). Further, in our case, it can be computed straightforwardly (it is instantaneous in comparison with the computation of the gradient).

A more physical justification for the use of this metric is that it corresponds to (approximate) imaginary time evolution \cite{hackl2020}, which converges exponentially fast for a gapped system and an expressive enough state manifold. Indeed, upon an infinitesimal imaginary time evolution $\upd \tau$ an RCMPS $\ket{x}$ evolves into $\ket{x} - \upd \tau H \ket{x}$. This latter state no longer belongs to the RCMPS manifold, and to get an approximate evolution we need to project down the evolution to the tangent space. More precisely, we want to find a direction $u\in \mathbb{R}^N$ in the tangent space such that $u^\mu \partial_\mu \ket{x} \simeq  - H \ket{x}$. It is obtained by minimizing $\|u^\mu \partial_\mu \ket{x} +  H \ket{x} \|^2$, which gives
\begin{equation}
    u^\mu= - [g^{-1}]^{\mu\nu}\partial_\nu (\bra{x}H\ket{x})\,
\end{equation}
provided $g_{\mu\nu}$ is invertible which we will assume here. This projected imaginary time evolution corresponds to the (imaginary) time dependent variational principle (TDVP) in the tensor network context \cite{vanderstraeten2019tangentspace}. In my opinion, the advantage of seeing imaginary TDPV simply as gradient descent with a different metric is that it makes it obvious the time step does not need to be infinitesimal, and can be chosen optimally with a line search.

In practice, I observed for $D\leq 4$ that quasi-Newton methods, which try to estimate the best metric (the Hessian) from the gradient at different iterations, are still reasonably efficient. For larger $D$, I found that the metric $g_{\mu\nu}$ becomes very singular near the ground state, which may explain why quasi-Newton methods fail to estimate the Hessian (which is likely very singular as well) and get stuck in plateaus. However, as we will see in \ref{sec:algorithm}, the tangent space approach I presented converges fast even for large values of $D$ as one would expect. For the optimization RCMPS in moderately large $D$, it is thus better to have an exact ``good'' metric, than an approximation of the best metric.

\subsection{Computing the metric}

\noindent The metric can be computed easily following~\cite{vanderstraeten2019tangentspace}. The first step is to define the tangent space vectors
\begin{align}\label{eq:tangent}
    \ket{V,W}_{Q,R} =\!\! \int\! \upd x \left[V_{\alpha\beta} \frac{\delta }{\delta Q_{\alpha\beta}(x)} + W_{\alpha\beta} \frac{\delta }{\delta R_{\alpha\beta}(x)}\right] \! \ket{Q,R} .
\end{align}
The complex matrices $V,W$ parameterize the direction in the tangent space, and $Q,R$ the point on the RCMPS manifold. We work in the translation invariant case, where $Q,R$ are taken position independent at the end, but the position argument $x$ in \eqref{eq:tangent} is necessary to know how operators are ordered. A crucial fact is that the tangent space is overparameterized and, with the left canonical choice \eqref{eq:left_gauge} we took for $Q$, \ie $Q = -iK -\frac{1}{2} R^\dagger R $, one is free to fix $V=-R^\dagger W$ without losing a linearly independent direction \cite{vanderstraeten2019tangentspace}. We may thus drop $V$ as a parameter, as it is fixed by $W$. With this choice, one can show \cite{vanderstraeten2019tangentspace} that the overlap between tangent vectors takes the particularly simple form
\begin{equation}\label{eq:overlap}
\begin{split}
    \langle W_1 | W_2 \rangle_{Q,R} &= \bra{\ell_0} W_2\otimes \bar{W}_1 \ket{r_0} \\
    &= \tr [W_2\rho_\text{ss} W_1^\dagger]
    \end{split}
\end{equation}
where $\rho_\text{ss}$ is the (normalized) stationary state of the Lindbladian $\mathcal{L}$ defined in \eqref{eq:def_Lindblad}.
We thus have $2D^2$ directions on the tangent space corresponding to the real and imaginary parts of the coefficients $W_{\alpha\beta}$. The metric is simply the bilinear map taking two $W$ and outputing the real part of \eqref{eq:overlap}. Note that the metric depends on the state only and is cheap to compute. Indeed, it does not require the resolution of ODEs which are the numerical bottleneck of RCMPS.

\subsection{Computing the gradient with an adjoint method}
\noindent To compute the gradient of the energy density in the $2D^2$ independent directions, one a priori needs $\propto 2D^2$ computations of expectation values each with a cost $\propto D^3$. However, using standard adjoint methods (a.k.a. backpropagation), one can compute the complete gradient with the same asymptotic cost as computing the energy, hence $\propto D^3$. In principle, this could be done by using a complex ODE solver compatible with automatic differentiation. In practice, because the ODEs involved have a special form, it is easy, efficient, and illuminating to implement the adjoint method directly. I illustrate the idea on the example of the computation of the gradient of a vertex operator, but it applies equally easily to the gradients of field monomials and kinetic term.

The expectation value of a vertex operator on a RCMPS is
\begin{equation}\label{eq:vertex_2}
    \langle V_b\rangle_{Q,R} = \tr\left\{\mathcal{P}\exp\left[\int_\mathbb{R} \upd x \, \mathcal{L} + b \mathcal{R}(x)\right]\cdot \rho_\text{ss}\right\}.
\end{equation}
Let us consider the gradient in the $W$ direction $\nabla_{W} \langle V_b\rangle_{Q,R}$ which is defined implicitly via
\begin{equation}
     \langle V_b\rangle_{Q+\varepsilon V,R + \varepsilon W} =  \langle V_b\rangle_{Q,R} + \varepsilon \nabla_{W} \langle V_b\rangle_{Q,R} + O(\varepsilon^2).
\end{equation}
Differentiating directly \eqref{eq:vertex_2} yields
\begin{equation}\label{eq:gradient_raw}
    \nabla_{W} \langle V_b\rangle=\!\!\int \!\!\upd y\, \tr\left\{ \mathcal{P}\e^{\int_{y}^{+\infty}  \!\!\mathcal{L}^b}\!\!\cdot\!\nabla_{W}\mathcal{L}^b(y)\!\cdot\mathcal{P}\e^{\int_{-\infty}^y\!\! \mathcal{L}^b}\!\!\!\cdot \rho_\text{ss}\right\}.
\end{equation}
with the notation $\mathcal{L}^b(x) = \mathcal{L} + b\mathcal{R}(x)$ and
\begin{equation}\label{eq:derivative_Lindblad}
\begin{split}
    \nabla_{W}\mathcal{L}^b(y) \cdot \rho = &V\rho + \rho V^\dagger + \frac{1}{2} \left(R\rho W^\dagger+W\rho R^\dagger\right) \\
    &+ b J(y) \left(W\rho + \rho W^\dagger\right)\, ,
    \end{split}
\end{equation}
recalling again that $V=-R^\dagger W$.
This gradient of $\mathcal{L}^b(y)$ appears in \eqref{eq:gradient_raw} between two evolution super-operators. Because a trace is taken at the end, we can replace the last part of the evolution from $y$ to $+\infty$ by its adjoint acting on the identity
\begin{equation}\label{eq:gradient_adjoint}
\begin{split}
    \nabla_{W} \langle V_b\rangle=\int \!\!\upd y\, \tr\bigg\{ &\left[\mathcal{P}\e^{\int_{y}^{+\infty}  \!\!\mathcal{L}^{b*}}\!\!\!\!\cdot \mathds{1}\right]  \\
    \times &\nabla_{W}\mathcal{L}^b(y)\!\cdot\left[\mathcal{P}\e^{\int_{-\infty}^y\!\! \mathcal{L}^b}\!\!\!\cdot \rho_\text{ss}\right]\bigg\}.
\end{split}
\end{equation}
where the adjoint $\mathcal{L}^{b*}(y)$ of $\mathcal{L}^b(y)$ is defined as
\begin{equation}
    \mathcal{L}^{b*}(y)\cdot \mathcal{O} = Q^\dagger\mathcal{O} + \mathcal{O} Q+ \frac{1}{2} R^\dagger \mathcal{O} R + b J(y) \left[R^\dagger \mathcal{O} + \mathcal{O}R \right].
\end{equation}
Writing as before $\rho_x=\mathcal{P}\e^{\int_{-\infty}^x\!\! \mathcal{L}^b}\!\!\!\cdot \rho_\text{ss}$ the solution of the forward problem and $\mathcal{O}_x=\mathcal{P}\e^{\int_{x}^{+\infty}  \!\!\mathcal{L}^{b*}}\!\!\!\!\cdot \mathds{1}$ the solution of the backward problem we get
\begin{equation}
    \nabla_{W} \langle V_b\rangle =\int \upd y \, \tr \left[\mathcal{O}_y \nabla_{W}\mathcal{L}^b(y)\cdot \rho_y\right]
\end{equation}
This can be further simplified exploiting the expression of $\nabla_{W}\mathcal{L}^b(y)$ \eqref{eq:derivative_Lindblad}, and one obtains all the components of the gradient from 2 matrices $M_W$ and $M_{W^\dagger}$
\begin{equation}\label{eq:gradient_compact}
    \nabla_{W} \langle V_b\rangle =\tr\left[M_W W + M_{W^\dagger} W^\dagger\right]
\end{equation}
where 
\begin{equation}\label{eq:matrices_integral} 
\begin{split}
    M_W&= \!\!\int \upd y\, -\!\rho_y \mathcal{O}_y R^\dagger + \frac{1}{2} \rho_y R^\dagger \mathcal{O}_y + b J(y) \rho_y \mathcal{O}_y \\
    M_{W^\dagger}&= \!\!\int \upd y\, -\! R\mathcal{O}_y\rho_y  + \frac{1}{2} \mathcal{O}_y R \rho_y + b J(y) \mathcal{O}_y \rho_y \,.
\end{split}
\end{equation}
In practice, one computes $\rho_y$ and $\mathcal{O}_y$ by solving the corresponding ODEs. The matrices $M_W$ and $M_{W^\dagger}$ are then obtained by evaluating the integrals in \eqref{eq:matrices_integral} with an efficient numerical method like the tanh-sinh quadrature~\cite{mori2001}. This gives an algorithm with a cost $\propto D^3$ to compute the full gradient of the expectation value of a vertex operator. The gradients of the kinetic term and of other potentials can be computed efficiently with the same method.

\subsection{Algorithm}\label{sec:algorithm}

\noindent We now have all the pieces to understand the optimization algorithm. The first step is to start from an initial guess. One can certainly do much smarter, but I started from uniformly random $R$ and $K$ matrices. This gives a very high starting energy density, but it fortunately decreases fast enough that initialization is a secondary concern for the bond dimensions I probed.

The second step is to compute the descent direction, obtained by acting with the inverse metric on the gradient. The gradient is computed with the backpropagation method described before, and is the most costly step, while the computation of the inverse metric is an immediate algebraic operation using \eqref{eq:overlap}.

The third step is to move $R$ and $Q$ in the descent direction. The step size need not be small, as in imaginary time evolution, and it is chosen to approximately yield the maximal energy decrease at each step. In practice, I used a backtracking line search with Armijo-Goldstein condition to find the the right amount to move at each step. Note that this is very similar to what was done by Ganahl \textit{et al.} in \cite{ganahl2017} for standard CMPS and the Lieb-Liniger model. Like them, I observed that this approach speeds up the optimization by roughly 2 orders of magnitude compared to imaginary time evolution with an optimal but fixed time step. Typically, results are converged after $\sim 10^2-10^4$ iterations depending on the coupling (convergence is slower near criticality), bond dimension (convergence is slower for large $D$), and random initial seed.

\section{Application to the self-interacting scalar}\label{sec:application}
\subsection{The model}
\noindent
To assess the soundness of RCMPSs, we consider the simplest non-trivial QFT in $1+1$ dimensions, the self-interacting scalar field with Hamiltonian 
\begin{equation}\label{eq:phi4Hamiltonian}
    H= :\left[\int_{\mathbb{R}} \frac{\pi^2}{2}+ \frac{(\partial_x \phi)^2}{2} + \frac{m^2}{2} \phi^2 + g\,  \phi^4 \,\right]: .
\end{equation}
The normal ordering is again done with respect to the creation and annihilation operators $a_k^\dagger,a_k$ which diagonalize the quadratic part of the Hamiltonian and are defined in \eqref{eq:modeexpansion}.

This model is a good case study because it is simple to define, even rigorously, as $H$ is a genuine renormalized Hamiltonian (self-adjoint, finite energy density). Yet, the model is not integrable, and carrying accurate computations out of the perturbative regime is non-trivial. The self-interacting scalar has been studied with a wide variety of methods: renormalized Hamiltonian truncation (without space-time discretization but finite size) \cite{rychkov2015}, infinite matrix product states (with space discretization but no finite size cutoff) \cite{milsted2013,vanhecke2019}, Monte-Carlo (with space-time discretization and finite size) \cite{schaich2009,bosetti2015,bronzin2019}, tensor network renormalization (with space-time discretization and finite size) \cite{kadoh2019,delcamp2020}, and, of course, (resummed) perturbative expansions (without cutoff, but perturbative)~\cite{serone2018}.

Out of these works, let us mention two that are particularly relevant for our study as they are carried in the Hamiltonian formalism. The study of Milsted \textit{et al.} \cite{milsted2013} is the the closest, in terms of method used, to what we shall do: the authors discretize the model in space, and find the ground state with translation invariant matrix product states, thus without IR cutoff, reaching unbeatable precision at the time. The drawback is the need to extrapolate the continuum limit, \ie the UV cutoff. On the other hand, the remarkably pedagogical Hamiltonian truncation study of Rychkov and Vitale \cite{rychkov2015} is carried directly in the continuum with the Hamiltonian \eqref{eq:phi4Hamiltonian}. In a nutshell, the authors introduce an IR and an energy cutoff that make the Hilbert space finite, and exactly diagonalize the resulting finite Hamiltonian matrix. In addition, they introduce a smart renormalization procedure which makes the convergence of the results faster as the cutoffs are lifted. The drawback is that the size of the Hilbert space (and thus the cost) is exponential in both cutoffs (I discuss it further in \ref{sec:discussion}). Further, while the renormalization procedure and careful extrapolations drastically improve precision, they destroy the variational nature of the results.

My objective is not so much to improve upon these studies in terms of raw numerical precision than in terms of conceptual simplicity, robustness, and scaling: with RCMPS one can in principle find the ground state of \eqref{eq:phi4Hamiltonian} directly in the continuum limit, without UV or IR cutoff, and in a variational way (that is, with rigorous energy upper bounds).
\subsection{Ground energy density}
\begin{figure}
    \centering
    \includegraphics[width=.99\columnwidth]{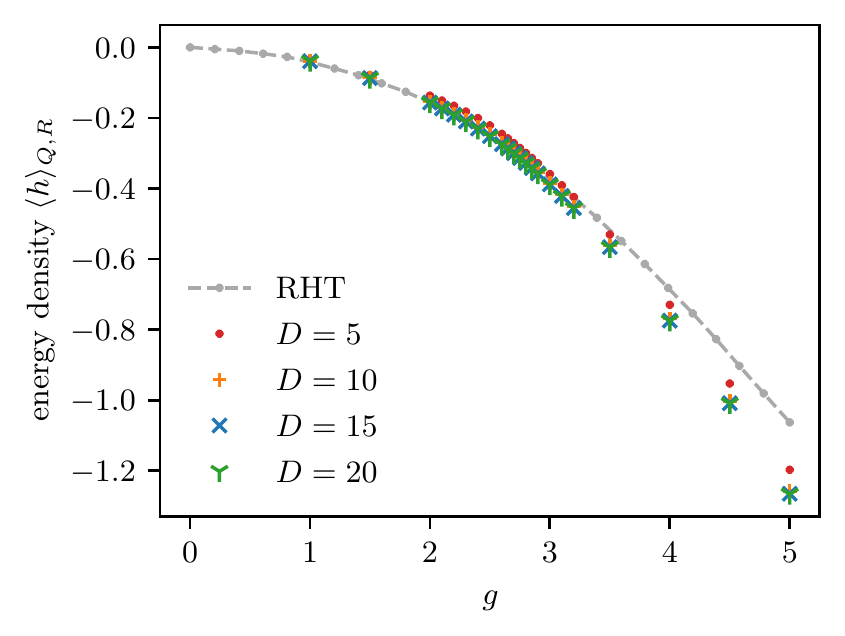}
    \caption{Ground state energy density of the $\phi ^4$ model as a function of the coupling $g$ for $m=1$. The RCMPS results are compared with the renormalized Hamiltonian truncation calculations from \cite{rychkov2015}, obtained with an IR cutoff $L=10$. Points are sampled more densely around the critical coupling $g_c\simeq 2.77$.}
    \label{fig:energy_density}
\end{figure}

\noindent 
The ground state energy density is finite and negative for the model we consider. This is clear given that the Fock vacuum for the free part $\ket{0}_a$, which is no longer the ground state for $g\neq 0$, already gives a zero energy because the Hamiltonian is normal ordered. Optimizing the RCMPS for $m=1$ indeed shows that the energy density is negative and decreases as the coupling is increased (see Fig. \ref{fig:energy_density}), at first quadratically in the coupling constant $g$ (as expected from perturbation theory \cite{rychkov2015}).

Importantly, the convergence in bond dimension is fast for all values of the coupling, even deep in the non-perturbative regime (which kicks in roughly at $g\geq 0.1$). As we can see in Fig. \ref{fig:energy_density} the energy as a function of $g$ is already qualitatively correct for $D=5$, and the points at larger bond dimensions ($D=10,15,20$) are essentially indistinguishable on this plot. At large coupling ($g \geq 3$ the results are substantially below those obtained with renormalized Hamiltonian truncation (RHT) \cite{rychkov2015}, which means they are more accurate since the method is truly variational. 

\begin{figure}
    \centering
    \includegraphics[width=0.99\columnwidth]{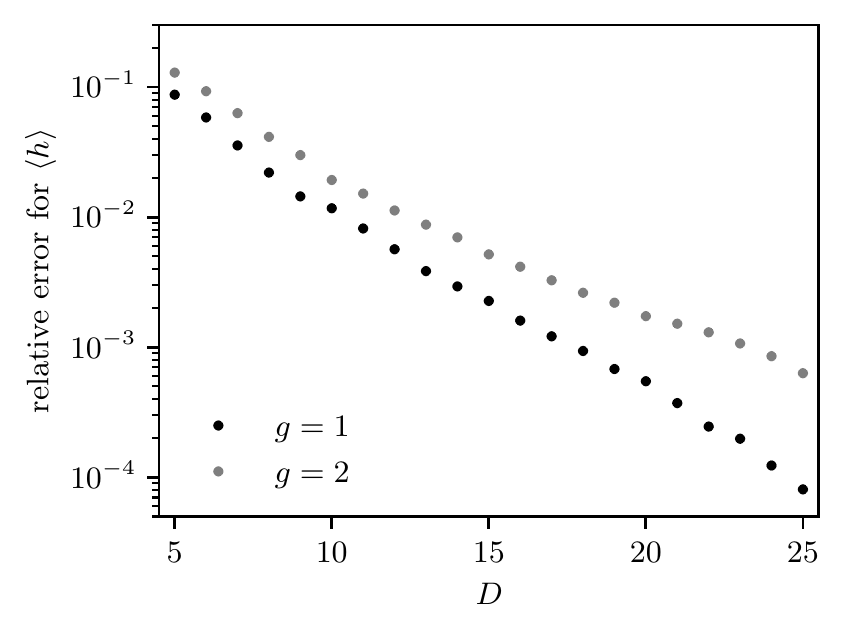}
    \caption{Relative error in the energy density as a function of the bond dimension $D$. The error is computed by using as reference the results at $D=32$: $\langle h \rangle_{g=1} \leq -0.0393547$ and $\langle h \rangle_{g=2} \leq -0.157214$. The latter should have a substantially lower error ($\simeq 10^{-5}$ and $\simeq 10^{-4}$ respectively) and thus be exact enough for the approximate computation of relative errors.}
    \label{fig:error}
\end{figure}

To estimate the error, it is not possible to compare with an exact solution ($\phi^4_2$ is not integrable), nor with earlier numerical results, \eg RHT, which are less precise even in their latest high precision development \cite{eliasmiro2017-2}. To get an accurate point of comparison, I simply considered a large $D$ estimate of the energy density as reference to estimate the error at lower bond dimensions. For $D=32$, I obtained the rigorous bounds $\langle h \rangle_{g=1} \leq -0.039354$ and $\langle h \rangle_{g=2} \leq -0.157214$, which can be used as fairly good estimates of the true values. The resulting errors for lower bond dimensions are shown in Fig. \ref{fig:error} for $g=1$ and $g=2$ and provide hints that the convergence to the ground energy is close to exponential in the bond dimension, or at least faster than any power law as one expects from the discrete \cite{huang2015computing}. Retrospectively, this fast convergence implies that our $D=32$ points of comparison are exact enough, with an error to the true ground state much smaller than that of the lower $D$ points whose error is estimated.

Note that the energy density obtained with RCMPS is the renormalized one and thus would be very difficult to estimate with a similar precision with lattice methods. Indeed, the latter give access to the total energy density, which diverges in the continuum limit. One would need to subtract the diverging tadpole part to get the finite renormalized contribution. As one gets closer to the continuum limit, obtaining this finite correction to a fixed precision requires a prohibitively high relative precision on the total energy.

Finally, note that a second order phase transition occurs for $g_c\simeq 2.77$ (for $m=1$) \cite{delcamp2020}. Just like Rychkov and Vitale~\cite{rychkov2015} with RHT, and as expected from the similarity with the 2d Ising model, we see no sign of this transition in the ground state energy density. As we will see, the transition appears more clearly once we look at observables like the expectation value of the field $\phi$ (which behaves like the Ising magnetization).

\subsection{Observables}
\begin{figure}
    \centering
    \includegraphics[width=0.99\columnwidth]{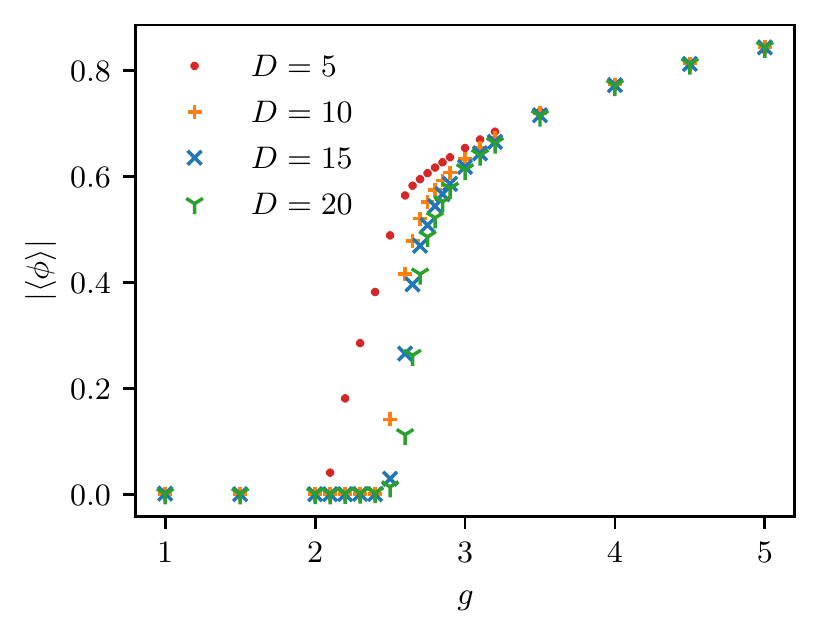}
    \caption{Absolute value of $\langle\phi \rangle$ taken in the approximate RCMPS ground state. The symmetry breaking as the coupling is increased is manifest. Away from the critical point, convergence is extremely fast as a function of the bond dimension.}
    \label{fig:phi1}
\end{figure}

\begin{figure}
    \centering
    \includegraphics[width=0.99\columnwidth]{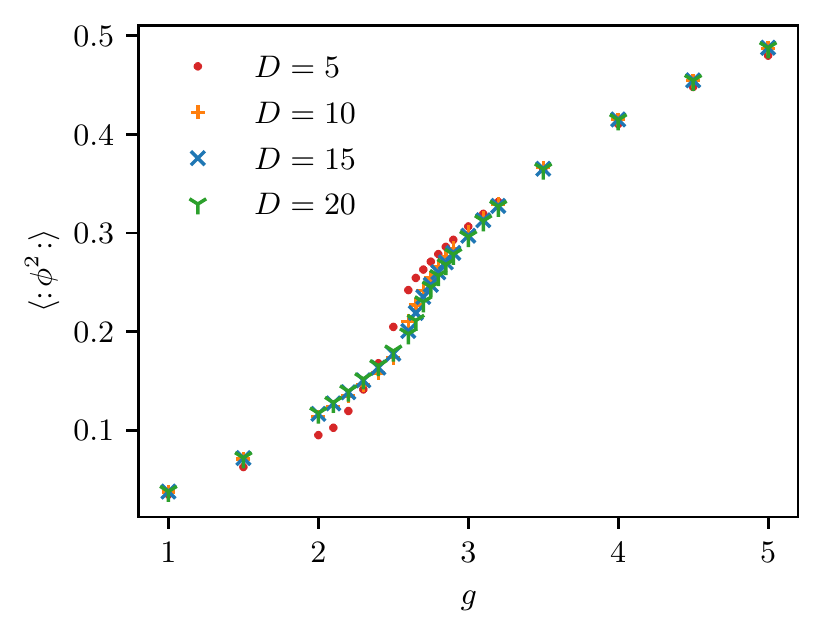}
    \caption{$\langle:\!\phi^2\!:\rangle$ in the approximate RCMPS ground state. The phase transition can also be seen, although less clearly than in Fig. \ref{fig:phi1}, through the divergence of the first derivative near criticality.}
    \label{fig:phi2}
\end{figure}

\noindent 
Once an RCMPS is optimized to approximate the ground state of a given Hamiltonian, expectation values of operators come essentially for free. As an illustration, I show the expectation values of $ \phi $ (Fig. \ref{fig:phi1}) and $:\!\phi^2\!:$ (Fig. \ref{fig:phi2}) but one could equally easily consider $:\!\phi^{42}\!:$ or $:\!\cosh(\phi)\!:$.

The expectation value of $\phi$, which is the equivalent of the Ising magnetization for the $\phi^4$ model, is the most instructive. It shows a clear spontaneous symmetry breaking around the expected coupling $g_c\simeq 2.77$.
To locate it more precisely (with a precision closer to the lattice extrapolations \cite{delcamp2020}) and estimate the critical exponents, one would need to compute more data points near the critical point, for a wide range of $D$, and use finite entanglement scaling techniques \cite{pollmann2009,stojevic2015,vanhecke2019}. This is left for future work. Consequently, the exceptional numerical accuracy of the ansatz is so far limited to the gapped phases on both sides of the critical point.

\section{Extensions} \label{sec:extensions}

\subsection{Adjustable characteristic length-scale}
The core idea of RCMPS compared to CMPS is to use creation operators such that the theory is exactly solved at short distance. We used the pair $a(x),a^\dagger(x)$ associated to the free part of the theory with mass $m$. As we discussed, this introduces a length-scale $m^{-1}$ for the exponentially decaying support of the Hamiltonian density, which is reminiscent of the lattice scale for lattice models. But this scale is arbitrary in our case and we could have chosen a different pair of creation-annihilation operators as long as the UV behavior is still exact. 

The simplest extension is to consider mass-adjustable RCMPS, that is states of the form
\begin{equation}
    \ket{Q,R,\tilde{m}} = \tr\left\{\mathcal{P} \exp\left[\int \upd x \, Q\otimes \mathds{1} + R\otimes a_{\tilde{m}}^\dagger(x)\right]\right\}\ket{0}_{\tilde{m}}
\end{equation}
and where $a_{\tilde{m}}$ are the operators diagonalizing the free theory of mass $\tilde{m}$, and $\ket{0}_{\tilde{m}}$ is the associated Fock vacuum. More explicitly, the field operators can be expanded in this new operator basis as
\begin{align}\label{eq:modeexpansion_alt}
    \phi(x) &= \frac{1}{2\pi} \int \upd k \sqrt{\frac{1}{2 \, \widetilde{\omega}_k}} \left(\e^{ikx} a_{\tilde{m},k} + \e^{-ikx} a^\dagger_{\tilde{m},k} \right) \\
        \pi(x) &= \frac{1}{2\pi} \int \upd k \sqrt{\frac{\widetilde{\omega}_k}{2}} \left(\e^{ikx} a_{\tilde{m},k} - \e^{-ikx} a^\dagger_{\tilde{m},k} \right) \, ,
\end{align}
where $\widetilde{\omega}(k) = \sqrt{k^2+\tilde{m}^2}$. At short distances, large $k$, the state $\ket{Q,R,\tilde{m}}$ still solves the theory exactly just like the RCMPS $\ket{Q,R}=\ket{Q,R,m}$, but the variable mass or inverse length-scale $\tilde{m} \neq m$ gives an additional degree of freedom one can optimize to better fit the IR. 

The computations with the mass variable RCMPS are more difficult than with the standard RCMPS, and the complications mainly come from the fact that the Hamiltonian density is not normal-ordered for the operators $a_{\tilde{m}},a^\dagger_{\tilde{m}}$. Evaluating the Hamiltonian density is thus tedious but doable, and I could carry the optimization. However, for the few values of coupling I tried, I obtained rather underwhelming results, only marginally improving the precision at the cost of a substantial increase in complexity and slower optimization.

A more thorough study should be done in future works. One could expect that carefully optimizing the length-scale $\tilde{m}^{-1}$ near criticality would yield more meaningful improvements, since the gap becomes much smaller than the mass $m$ appearing in the Hamiltonian (corresponding to the one-loop renormalized mass). A more systematic exploration of Bogoliubov transformations would be interesting as well. In principle, one can change the $a,a^\dagger$ into new operators $b,b^\dagger$ by replacing $\omega_k$ in the mode expansion \eqref{eq:modeexpansion} with any function $\Omega_k$ with the same large $k$ behavior so that expectation values are still UV finite. This should provide a substantial gain in expressiveness at fixed bond dimension, but the computations would be more involved.

Finally, I would like to emphasize that this idea of non-local change of basis to make the continuum limit well behaved or even simply increase expressiveness, which comes at the cost of making the Hamiltonian exponentially decaying instead of local, could be used on the lattice as well.

\subsection{Excitation spectrum and beyond}
\noindent
Once optimized, a RCMPS can be used to compute all correlation functions at equal time for no additional minimization cost. As I argued, this also gives some dynamical information because of Lorentz invariance, but can one get more? 

In principle, one can use TDVP in real-time to evolve states which means we have access to all dynamical properties. Note however in that case that one cannot use the trick of taking optimally large time steps, since, in real time, we no longer have a global minimization problem. 

Using standard tangent-space CMPS techniques, one also has access to the excitation spectrum \cite{vanderstraeten2019tangentspace}. In a nutshell, the idea is to diagonalize the Hamiltonian in the tangent space $\ket{V,W}_{Q,R}$ which is a vector space orthogonal to the ground state (in the gauge we chose). This would give approximations only to the excited states with zero momentum, but simple local modifications allow to target states of non-zero momenta as well \cite{vanderstraeten2019tangentspace}. Carrying such computations requires evaluating matrix elements of the form $\bra{V',W'} h \ket{V,W}_{Q,R}$, which should be doable. A comparison of this spectral data from the one one could extract from the two-point function would be interesting.

\subsection{Other quantum field theories}
\noindent 
I illustrated the use of RCMPS on the self-interacting scalar field only, but it can in principle be applied to almost all theories in $1+1$ dimensions. Other bosonic theories with polynomial interactions (\eg $:\!\phi^6\! :$) can be treated without new technique. Scalar theories with exponential potentials, like the Sine-Gordon and Sinh-Gordon models can also be dealt with immediately since expectation values of vertex operators are straightforward to compute. 

Fermionic theories could also be dealt with directly, with a minor subtlety related to regularity conditions \cite{haegeman2013}, that is conditions on $R,Q$ one has to impose to make expectation values finite. The Gross-Neveu model \cite{gross1974}, which has already been studied with various UV cutoffs with tensor networks \cite{haegeman2010-relativistic,roose2020lattice}, would be an interesting candidate to probe the behavior of a ``just renormalizable'' theory. Alternatively, a large class of Fermionic models, like the Thirring model, can already be dealt using bosonization and the results of the present paper.

There are however serious difficulties remaining to extend the method to relativistic QFT in $2+1$ and $3+1$ dimensions. The first is specific to the Hamiltonian formalism, as renormalized Hamiltonians are more difficult to define in higher dimensions: normal ordering is no longer sufficient and the renormalized Hamiltonian no longer acts on the free Fock space \cite{glimm1968}. Nonetheless, recent progress was made using renormalized Hamiltonian truncation \cite{elias-miro2020} and lightcone conformal truncation \cite{anand2020nonperturbative}, with promising numerical results, showing that the Hamiltonian approach is still reasonable in $2+1$ dimensions and in the continuum. 

The second difficulty is related to continuous tensor network states themselves, that is the higher dimension equivalent of CMPS. In the non-relativistic setting, these states have been proposed in~\cite{tilloy2019}, but evaluating correlation functions is so far efficient only for Gaussian states~\cite{karanikolaou2020gaussian}. Indeed, the equivalent of the finite transfer matrix $\mathbb{T}$ of CMPS in $2+1$ dimensions is an operator acting on (two copies of) the Fock space of a $1+1$ dimensional relativistic QFT. In principle, one could bootstrap the present approach and evaluate correlation functions (or the energy density) in $2+1$ using a boundary RCMPS approach. In a nutshell, one would find the stationary state of the transfer matrix as a (large bond dimension) RCMPS. Every evaluation of an expectation value in $2+1$ would be done at the cost of a full optimization in $1+1$. This is the typical dimensional reduction obtained with tensor network approaches, where a physical dimension is traded for a variational optimization. Clearly, optimizing RCMPS is so far orders of magnitudes too slow to be used as a routine to be called many times in a full optimization process. I hope the present work can stimulate work in this direction. 

\section{Discussion} \label{sec:discussion}
\noindent
In this paper, I have presented a new class of states, the relativistic continuous matrix product states, that is adapted to relativistic quantum field theories. It is usable directly in the thermodynamic limit (no IR cutoff) and is exact at short distances, all the way down (no UV cutoff). The bond dimension $D$ controls the expressiveness of this class, and a state has $2D^2$ independent parameters. 

The comparison between RCMPS and Hamiltonian truncation is a good illustration of the difference between linear and non-linear methods. Hamiltonian truncation (up to its renormalization refinements) is a linear approach. The candidate ground state is expanded in a truncated Hilbert space. The energy is quadratic in the coefficients, and thus minimized efficiently as a linear problem. The price to pay is a lack of extensivity, the size of the Hilbert space grows exponentially as the size of the system increases for a fixed energy cutoff (and thus fixed precision). A side effect is that the number of parameters needs to be huge to reach good precision, with the candidate ground state written as a linear superposition of typically $10^4$ to $10^6$ states \cite{eliasmiro2017-2}. In contrast, using a continuous tensor network ansatz as we did, we work with a manifold of states, and the energy density is a highly non-linear function to optimize. Nonetheless, it can still be done efficiently as I showed in \ref{sec:optimization}. Further, having a non-linear class of states allows us to have an extensive ansatz, where going to the thermodynamic limit does not increase the number of parameters (in the translation invariant case). The results at $D=5$ provide an approximation to the ground state with only $2D^2=50$ independent real parameters which is already more accurate than RHT at strong coupling. 

This parsimonious encoding of the state translates into a better asymptotic behavior of the approximation. The error of Hamiltonian truncation decreases as a power law in the truncation energy $E_T$, while the cost is exponential in $E_T$ (see \eg \cite{eliasmiro2017-2}). In contrast, for RCMPS, the cost of the optimization is polynomial in $D$ while, at least for the model I considered, the error seems to converge exponentially fast to zero as a function of $D$. Rigorous results for MPS \cite{huang2015computing} lead one to expect a convergence at least faster than any power law.  This favorable scaling should be explored with more careful numerical simulations, and it would be interesting to see if it can be proved rigorously. Nonetheless, even without such a proof, RCMPS already provide rigorous and rather tight energy upper bounds. These rigorous results are made possible because we work directly in the continuum, without the need for extrapolations. 

Naturally, there is a lot of work to do to generalize the RCMPS to a wider a range of quantum field theories. But I hope the present paper will motivate others to pursue this necessary exploration.

\begin{acknowledgments}
\noindent I am grateful to have had discussions with Patrick Emonts, Tommaso Guaita, and Teresa Karanikolaou. They helped me realize the subtleties of minimization on a manifold, which was crucial for this work to succeed. I also thank Ignacio Cirac for helpful comments and for his support. Finally, I thank Jutho Haegeman, Karel Van Acoleyen, and Frank Verstraete, for helpful comments on an earlier version.
\end{acknowledgments}

\bibliography{main}

\end{document}